# Computational Protein Design with Deep Learning Neural Networks

Jingxue Wang[1], Huali Cao[1], John Z.H. Zhang[1-4], and Yifei Qi[1,2*]

[1]*Shanghai Engineering Research Center of Molecular Therapeutics and New Drug Development, School of Chemistry and Molecular Engineering, East China Normal University, Shanghai, 200062, China*
[2]*NYU-ECNU Center for Computational Chemistry at NYU Shanghai, Shanghai 200062, China*
[3]*Department of Chemistry, New York University, NY, NY 10003, USA*
[4]*Collaborative Innovation Center of Extreme Optics, Shanxi University, Taiyuan, Shanxi 030006*

[*]Correspondence to: yfqi@chem.ecnu.edu.cn

## Abstract
Computational protein design has a wide variety of applications. Despite its remarkable success, designing a protein for a given structure and function is still a challenging task. On the other hand, the number of solved protein structures is rapidly increasing while the number of unique protein folds has reached a steady number, suggesting more structural information is being accumulated on each fold. Deep learning neural network is a powerful method to learn such big data set and has shown superior performance in many machine learning fields. In this study, we applied the deep learning neural network approach to computational protein design for predicting the probability of 20 natural amino acids on each residue in a protein. A large set of protein structures was collected and a multi-layer neural network was constructed. A number of structural properties were extracted as input features and the best network achieved an accuracy of 38.3%. Using the network output as residue type restraints was able to improve the average sequence identity in designing three natural proteins using Rosetta. Moreover, the predictions from our network show ~3% higher sequence identity than a previous method. Results from this study may benefit further development of computational protein design methods.

# Introduction

Proteins perform a vast number of functions in cells including signal transduction, DNA replication, catalyzing reactions, etc. Engineering and designing proteins for specific structure and function not only deepen our understanding of the protein sequence-structure relationship, but also have wide applications in chemistry, biology and medicine.[1] Over the past three decades, remarkable successes have been achieved in protein design, in which some of the designs were guided by computational methods. Examples of some recent successful computational protein designs include novel folds,[2] novel enzymes,[3,4] vaccines,[5,6] antibodies,[5,7,8] novel protein assemblies,[9-13] ligand-binding proteins,[14,15] and membrane proteins.[16-18] Comprehensive coverage of the protein designs until 2014 is provided by Samish,[19] and more recent ones are reviewed elsewhere.[20-23] In general, the input of computational protein design is the backbone structure of a target protein (or part of a target protein). Through computational sampling and optimization, sequences that are likely fold to the desired structure are generated for experimental verification. The scoring function usually contains physics-based terms such as van der Waals and electrostatic energy as well as knowledge-based terms such as sidechain rotamer[24] and backbone dihedral preference obtained from statistics of protein structures.[25,26] In many cases, the sequences from computational design are subject to further filtering by considering various factors such as shape complementarity[9] and *in silico* folding free energy landscape,[27,28] where human experience and familiarity with the designed protein play an important role, indicating that there is a gap between current design methods and fully automatic designs.

On the other hand, while the number of known protein structures is increasing rapidly, the number of unique protein folds is saturating. As of July 2017, there are ~132,000 structures in the protein data bank (PDB)[29] with a yearly increase of ~10,000, but the number of unique folds has not changed in the past few years, suggesting more data are accumulated on each fold, and therefore statistical learning and utilizing the existing structures are likely able to improve the design methods.[30,31] Recently, two statistical potentials for protein design have been developed,[32,33] and the ABACUS potential[34] has been successfully used in designing proteins.[33,35] While these statistical potentials have a physical basis, machine learning especially deep-learning neural network has recently become a popular method to analyze big data sets, extract complex features, and make accurate predictions.[36]

Deep-learning neural network, as a machine learning technique, is becoming increasingly powerful with the development of new algorithms and computer hardware, and has been applied to learning massive data sets in a variety of fields such as image recognition,[37] language processing,[38] and game playing.[39] Particularly in computational biology/chemistry, it has been used in protein-ligand scoring,[40-42] protein-protein interaction prediction,[43] protein secondary structure prediction,[44-49] protein contact map prediction,[50-52] and compound toxicity[53,54] and liver injury prediction,[55] among others.[56] In many cases it shows better performance than other machine learning methods. The advantage of using deep neural network is that it can learn high-order features from simple input data such as atom coordinates and types. The technical details such as network architecture, data representations vary from application to application, but the fundamental requirement of applying deep neural network is the availability of a large amount of data. With the aforementioned rich protein structure data available, it is promising to apply deep neural network in computational protein design. Zhou and coworkers have used the neural network approach to tackle this problem and developed the SPIN method to predict the sequence profile of a protein given the backbone structure.[57] The input features of SPIN include φ, ψ dihedrals of the target residue, sequence profile of 5-residue fragment derived from similar structures (the target residue and four subsequent residues), and a rotamer-based energy profile of the target residue using the DFIRE potential.[58] SPIN was trained on 1532 non-redundant proteins and reaches a sequence identity of 30.3% on a test set containing 500 proteins. This sequence identity is at the lower boundary of homologous protein[59] and is not sufficient to improve protein design significantly.

In this study, we applied deep-learning neural networks in computational protein design using new structural features, new network architecture, and a larger protein structure data set, with the aim of improving the accuracy in protein design. Instead of taking the whole input structure into account, we use a sliding widow method that has been used in protein secondary structure prediction, and predict the residue identity of each position one by one. We consider the target residue and its neighboring residues in three-dimensional spaces, with the assumption that the identity of the target residue should be compatible with its surrounding residues. We collected a large set of high-resolution protein structures and extracted the coordinates of each residue and its environment. The performance of the neural network on different input setups was compared, and application of the network outputs in protein design was investigated.

# Results

**Network architecture, input, and training**
The input of the computational protein design problem is the backbone structure of a protein (or part of a protein). Instead of predicting the residue types of all positions in the input protein simultaneously, we consider each target residue and its neighbor residues (for simplicity, non-protein residues are not considered). In the simplest case, we consider a target position and its closest neighboring residue determined by $C_\alpha$-$C_\alpha$ distance, and feed their input features to a neural network that consists of an input layer, several hidden layers and a softmax layer as output. The output dimension of the softmax layer is set to 20 so that the 20 output numbers that sum to one can be interpreted as the probabilities of 20 residue types of the target residue. This network is named residue probability network hereinafter (**Figure 1A**). Such a simple network that considers only one neighbor residue obviously cannot make satisfactory predictions. In this study, we take into account the target residue and its 10-30 neighbor residues by repeatedly using the residue probability network that shares the same parameters. This setup is similar to the application of convolution layer in image recognition where the same convolution network is applied to different regions of the input image. One drawback of this setup is that the output of each target-neighbor residue pair is equally weighted. Apparently, some neighbor residues have larger impacts on the identity of the target residue than others. To overcome this, we construct another network that takes the same input as the residue probability network but outputs a single number as the weight (**Figure 1B**). The output of the residue probability network is multiplied by this weight and then concatenated. Several fully-connected layers are then constructed on top of the weighted residue probabilities and a 20-dimentional softmax layer is used as the final output (**Figure 1C**), which can be interpreted as the probabilities of 20 residue types of the target residue.

The input for the residue probability and weight network consists of features from the target residue and one of its neighbor residues (**Figure 1C**). The features include basic geometric and structural properties of the residues such as $C_\alpha$-$C_\alpha$ distance, *cos* and *sin* values of backbone dihedrals φ, ψ and ω, relative location of the neighbor residue to the target residue determined by a unit vector from the $C_\alpha$ atom of the central residue to the $C_\alpha$ atom of the neighbor residue, three-type secondary structures, number of backbone-backbone hydrogen-bonds, and solvent accessible surface area of backbone atoms (see Methods). To train the neural network, we collected high-resolution protein structures from PDB using filtering conditions including structure determination method, resolution, chain length, and sequence identity (see Methods). Briefly, three data sets are prepared based on three sequence identity cutoffs (30%, 50%, and 90%, referred to as SI30, SI50, and SI90) to remove homologous proteins. For each of these data sets, each residue and its *N* (*N*=10, 15, 20, 25, 30) closest neighbor residues based on $C_\alpha$-$C_\alpha$ distance are extracted as a cluster. These clusters are randomly split into five sets for five-fold cross-validation. Hereinafter, we will use SI30N10 to refer to the dataset from 30% sequence identity cutoff with 10 neighbor residues. Similar naming rules apply to other datasets as well. The number of layers and nodes in each fully-connected layer were determined by training and test on the smallest data set SI30N10. The neural network training was performed for 1000 epochs to ensure convergence.

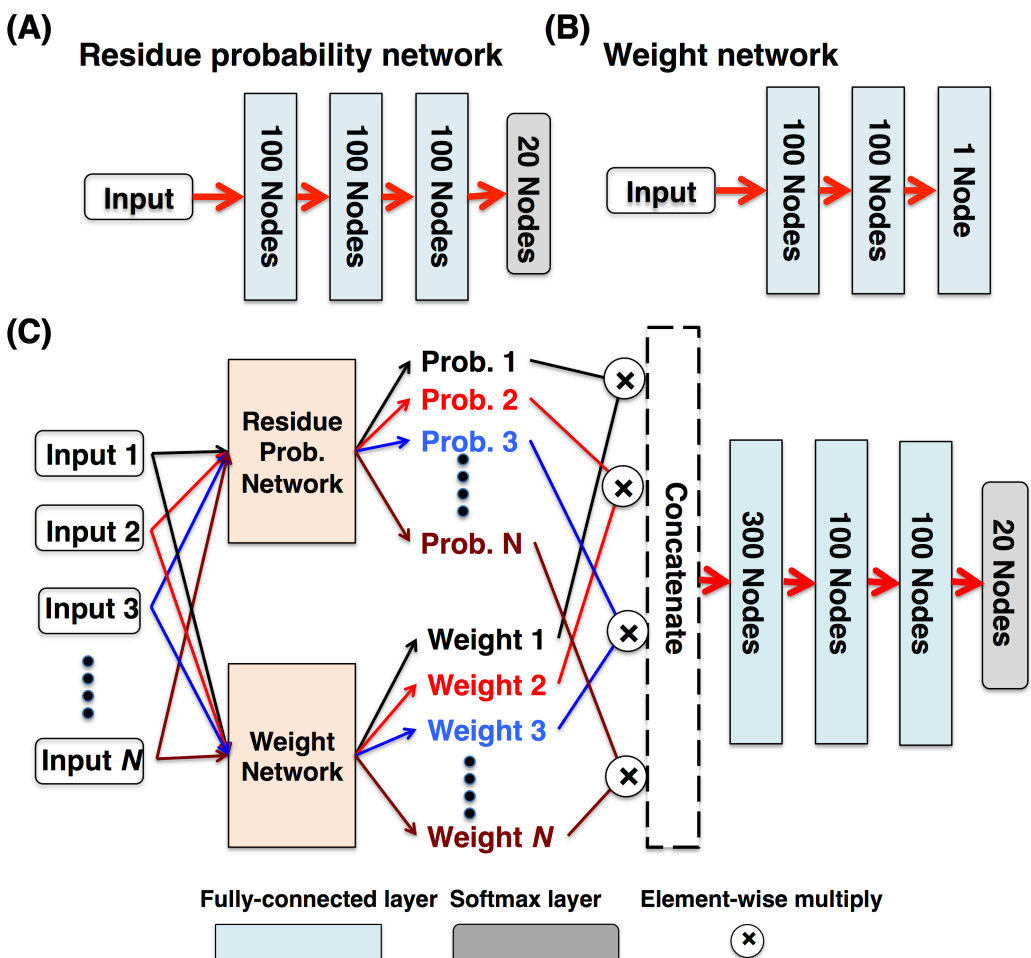

Figure 1. Architecture of the neural networks. (A) The residue probability network, (B) Weight network, and (C) The full network. The residue probability and weight networks are used as subnetworks that share the same set of network parameters for different inputs. Each *input* consists of the features from the target residue and one of its neighbor residues.

**Overall and amino acid specific accuracy**
**Table 1** shows the overall accuracy (percent of residues that are correctly predicted) and standard deviations on different datasets from five-fold cross-validation. As expected, datasets with higher protein identity cutoffs show better accuracy due to more data samples and higher similarities between samples. However, considering that the number of data samples almost doubled from SI30 to SI90 dataset, the improvement in accuracy is not significant. Furthermore, in each protein identity cutoff, including 15 neighbor residues show best accuracy. Including fewer neighboring residues likely under-represents the environment of the target residue, whereas including too many neighbor residues will generate noises in the inputs and thus require more data samples for training. An alternative way of extracting the neighbor residues is to use a certain distance cutoff. However, this strategy requires that the input size of the neutral network to be flexible, which will be investigated in future studies.

Table 1. Accuracy from five-fold cross-validation of the neural network on different datasets with different number of neighbor residues.

| Identity cutoff | N=10 | N=15 | N=20 | N=25 | N=30 |
|---|---|---|---|---|---|
| 30% | 0.329 (0.001)* | **0.340 (0.005)** | 0.333 (0.009) | 0.331 (0.006) | 0.321 (0.015) |
| 50% | 0.353 (0.003) | **0.364 (0.005)** | 0.358 (0.005) | 0.359 (0.006) | 0.342 (0.007) |
| 90% | 0.367 (0.001) | **0.383 (0.004)** | 0.382 (0.006) | 0.379 (0.007) | 0.352 (0.013) |

*Numbers in parentheses are standard deviations.

We next exam the amino-acid specific accuracy using the results from the SI90N15 dataset that has the best overall accuracy. To this end, we define the recall and precision for each amino acid. Recall is the percent of native residues that are correctly predicted (recovered), and precision is the percent of predictions that are correct. Pro and Gly have higher recall and precision than other residues with Pro achieves 92.1% recall and 62.7% precision (**Figure 2**). This is because Pro has an exceptional conformational rigidity and Gly is highly flexible in terms of backbone dihedrals. A neural network can easily learn these distinct structural properties. The amino acids that have lower recall/precision generally have lower abundance in the training set, for example Met, Gln and His, although we already applied bias to these low-abundance amino acids in training. To further characterize the amino-acid specific accuracy, we calculated the probability of each native amino acid being predicted as 20 amino acids, and plot it in a 2D native *vs* predicted heat map (**Figure 3**). The amino acids in x- and y-axis are ordered by their properties and similarities with each other. The diagonal grids show higher probabilities, as expected. Interestingly, there are several groups along the diagonal including RK, DN, VI, and FYW, indicating that the neural network frequently predicts one amino acid as another within each group. Considering the similarities of amino acids within each group, replacing one amino acid with another from the same group probably does not disrupt the protein structure, which suggests that the neural network may mispredict the native amino acid, but still provide a reasonable answer.

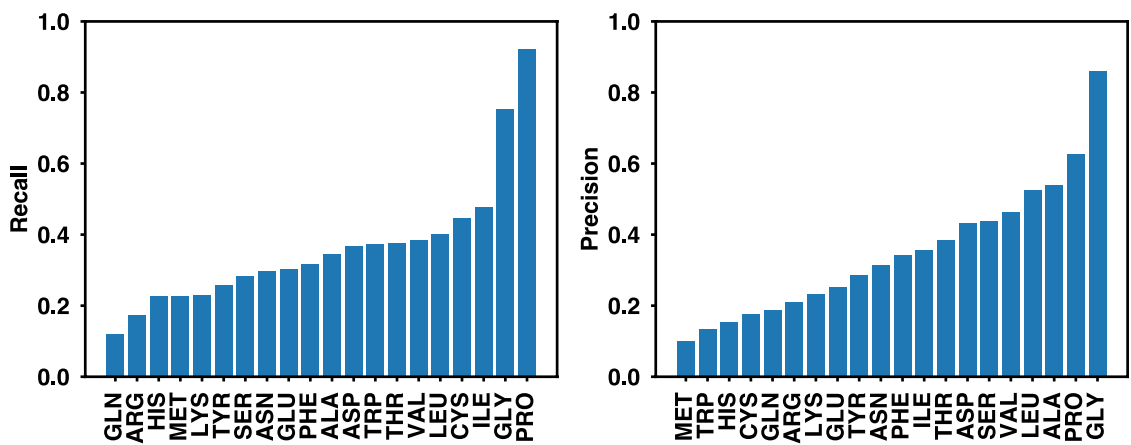

Figure 2. Recall and precision of different amino acids of the network trained on the SI90N15 dataset. Recall is the percent of native residues that are correctly predicted (recovered), and precision is the percent of predictions that are correct.

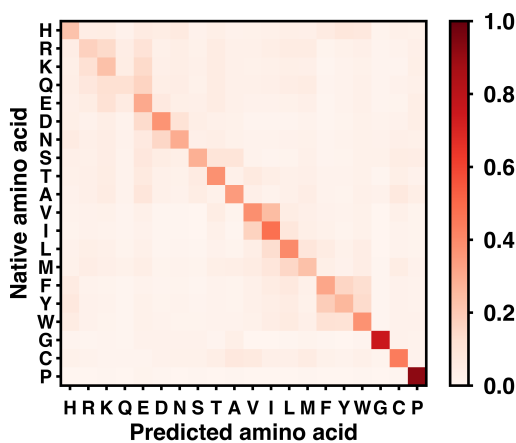

Figure 3. Probability of each amino acid being predicted as 20 amino acids.

**Top-*K* accuracy and its application in protein design**

Because the output of the neural network is the probabilities of 20 amino acids at a target position, in addition to the accuracy mentioned above, it is also possible to calculate the top-*K* accuracy: if the native amino acid is within the top-*K* predictions (*K* amino acids that have the highest probabilities), the prediction is considered correct. The top-2, 3, 5, and 10 accuracy of the network trained on the SI90N15 dataset reaches 54.3%, 64.0%, 76.3%, and 91.7% respectively, suggesting the native amino acids are enriched in the first half of the predictions (**Figure 4**). A simple application of such information is to restrain the available amino acid types at a target position during protein design. As an illustrative example, we applied the top-3, 5, and 10 predictions as residue-type restraints in designing three proteins including an all-α protein (PDB ID 2B8I[60]), an all-β protein (PDB ID 1HOE[61]), and a mixed αβ protein (PDB ID 2IGD, **Figure 5**). None of these proteins are included in our training set. The crystal structures of these proteins were used as inputs for the neural network trained on SI90N15 dataset. The top-3, 5, and 10 amino acids for each position were used as restraints in the fixed-backbone design program *fixbb* in Rosetta.[62] As a control, we listed the top one accuracy of the neural network on these proteins, and also performed fixed-backbone design without any residue-type restraints (all 20 natural amino acids are allowed at each position). As *fixbb* uses a stochastic design algorithm, we generated 500 sequences for each protein and calculated the average sequence identity to the native proteins (**Table 2**). In the three proteins, using information from the neural network predictions improves the average sequence identity, but the best *K* value is system dependent, and in some cases the results are worse than those in restraints-free designs (e.g., top-1 in 1HOE).

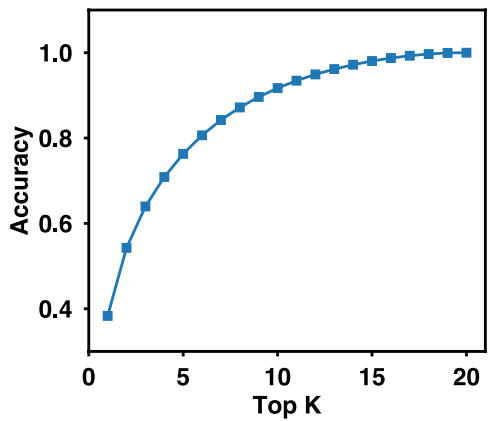

Figure 4. Top-*K* accuracy of the neural network trained on the SI90N15 dataset.

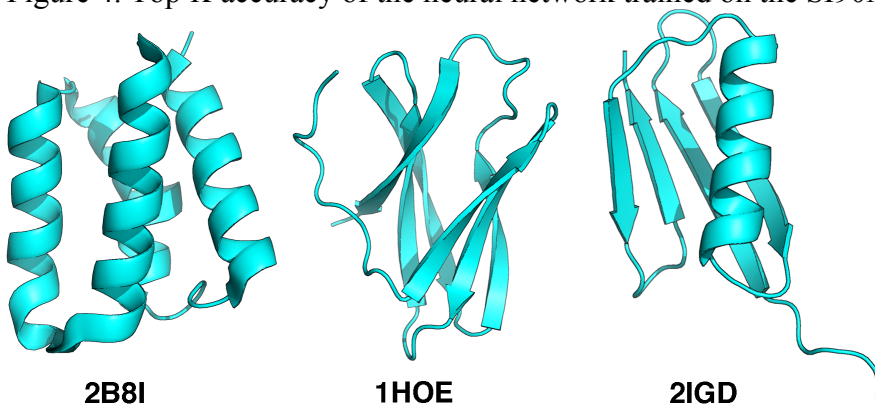

Figure 5. Structures of the proteins used in protein design with residue-type restraints.

Table 2. Average sequence identity of Rosetta fixed-backbone design on three proteins with/without residue-type restraints.

| Protein | No-restrain[*] | Top 1 | Top 3[*] | Top 5[*] | Top 10[*] |
|---|---|---|---|---|---|
| 2B8I | 0.276±0.033 | 0.337 | 0.306±0.017 (0.558) | **0.354±0.021 (0.688)** | 0.293±0.037 (0.883) |
| 1HOE | 0.408±0.026 | 0.338 | **0.473±0.018 (0.635)** | 0.441±0.018 (0.689) | 0.416±0.028 (0.851) |
| 2IGD | 0.409±0.034 | **0.475** | 0.473±0.023 (0.705) | 0.401±0.028 (0.754) | 0.408±0.032 (0.967) |

[*]Sequence identities are presented as average ± standard deviation from 500 designs. Numbers in parentheses are maximal possible identities given the residue-type restraints.

**Comparison with SPIN**

Finally, we compare the performance of our network with SPIN developed by Zhou and coworkers.[57] SPIN was trained on 1532 non-redundant proteins and reaches a sequence identity of 30.3% on a test set containing 500 proteins. The training and test set were collected using a sequence identity cutoff of 30%. SPIN was also evaluated on 50 proteins from the 500 test proteins for comparison with Rosetta.[62] As the structure ID of these 50 proteins are known, we set out to compare out network with SPIN on these 50 proteins. For a fair comparison, we re-trained our network on the SI30N15 dataset without the 50 proteins. **Table 3** lists the average sequence identity from both methods when top 1 to 10 predictions are considered. Our network shows ~3% higher identity than SPIN. The number of data samples almost tripled in our study (~1.5 million training residues for our network and ~0.45 million residues for SPIN assuming each protein has 300 residues) but the improvement of accuracy is not significant, indicating certain limitation in learning sequence information from protein structures.

Nonetheless, the networks trained on the larger data set in this study could still be beneficial to computational protein design. As in real applications, amino acid probability learned on larger data set could be more useful, as long as it is not biased. As an example, we tested both methods on the *de novo* designed protein Top7 (PDB ID 1QYS,[2] not included in our training set). The top 1 prediction from SPIN shows an identity of 0.250, while the top 1 predictions from the SI30N15, SI50N15, and SI90N15 network have identities of 0.283, 0.304, and 0.402.

In addition to comparing sequence identities, we also compared out predictions with the position-specific scoring matrix (PSSM) from PSI-BLAST.[63] The PSSMs of the 50 test proteins were obtained by running PSI-BLAST against the non-redundant protein sequences database available at ftp://ftp.ncbi.nlm.nih.gov/blast/db/, and converted to pseudo probability matrixes of 20 amino acids at each residue. The root mean square error (RMSE) of the matrixes to those predicted by our network and SPIN were calculated. Our network and SPIN show very similar RMSE values (0.139 for our network and 0.141 for SPIN). It should be noted that SPIN was trained on PSSMs from PSIBLAST for predicting sequence profiles whereas our network was trained on protein sequences only.

Table 3. Average sequence identity of SPIN and our network on 50 test proteins.

|  | Top 1 | Top 2 | Top 3 | Top 5 | Top 10 |
|---|---|---|---|---|---|
| SPIN | 0.302 | 0.453 | 0.552 | 0.677 | 0.868 |
| This study[*] | 0.330 | 0.487 | 0.585 | 0.717 | 0.896 |
|  | (0.002) | (0.005) | (0.002) | (0.001) | (0.002) |

[*]Numbers in parentheses are standard deviations from 5 networks trained on the same dataset with different random number seeds.

## Discussions

In this study, we have developed deep-learning neural networks for computational protein design. The networks achieve an accuracy of 38.3% on the dataset with 90% sequence identity cutoff when 15 neighboring residues are included. This accuracy is limited not only by our neural network approach but also by the nature of protein structures. It is known that two proteins with low sequence identity (~30%) can fold into similar structures.[59] In a DNA repair enzyme 3-methyladenine DNA glycosylase, the probability that a random mutation can inactivate the protein was found to be 34%±6%, indicating a large proportion of mutations can be tolerated in this protein.[64] Moreover, residues at active sites are subject to functional restraints and are not necessary the most stable ones.[65] Our neural network approach is similar to a structural comparison method that extracts and integrates similar local structures from known structures. Therefore, its accuracy is limited by the average number of tolerable amino acids at each residue in the training set. Fortunately, the native amino acid is concentrated in the top predictions (top-5 and 10 accuracies are 76.3% and 91.7%). By integration the network output with molecular-mechanics scoring functions, it should be possible to identify the correct amino acid from the top predictions and further improve the accuracy. Particularly, the network preforms well on Gly and Pro, due to its ability to learn distinct structural features, but less satisfying on hydrophobic residues that are likely more important for the correct folding of a protein. Including solvation energy in the molecular mechanics scoring functions is probably a promising way for future development.

In our approach, the environment of a target residue is simply considered using the $N$ closest residues based on $C_\alpha$-$C_\alpha$ distances. This method may exclude some residues that have important interactions with the target residue. To quantitatively characterize this, we calculated the distance rank ($M$, the rank of $C_\alpha$-$C_\alpha$ distance of a neighbor residue among all residues surrounding the target residue) of neighbor residues that have contacts (heavy atom distance < 4.5 Å) with the target residue in our dataset, and found that 96.2% contacting residues have $M\leq20$, and 98.9% contacting residues have $M\leq30$, which means 3.8% and 1.1% of the contacting residues are not included in the environment if $N$=20 and 30, respectively. Moreover, for terminal residues that are highly exposed, 20 neighbors may contain residues that do not have contacts with the target residue, which will generate noises in the inputs. Using distance cutoff instead of residue number cutoff may solve this problem. However, the distance cutoff method requires the input size to be highly flexible from several residues to tens of residues, which should be carefully considered during network construction.

Knowing the possible amino acids with good confidence at the designing positions may reduce the search space significantly and increase the chance to make a successful design. Our test of Rosetta design on three proteins shows that it is possible to improve the sequence identity by using the output from our neural network as residue-type restraints. However, the optimal number of amino acids to be used as restraints is system dependent. More importantly, in our neural network, the prediction on each residue is independent from each other. For real designs, it is important to simultaneously consider the identities of the neighbor residues by using molecular-mechanics-based or statistical scoring functions like the ones in Rosetta. In this regard, the predicted probability of each amino acid should be explicitly taken into account. As the prediction of a trained neural network on a protein structure only takes several seconds, we expect our approach to pave the way for further development of computational protein design methods.

## Methods
### Datasets and input features
The training set was collected from PDB[29] using the following criteria: (1) the structure is determined with x-ray crystallography, (2) the resolution is better than 2 Å, (3) the chain length is longer than 50, and (4) the structure does not have any DNA/RNA molecules. To investigate the effects of sequence homology on prediction accuracy, the structures that satisfy these conditions were retrieved with 30%, 50%, and 90% sequence identities. The resulting entries were cross-referenced with the OPM database[66] to remove membrane proteins. Structures that have D-amino acids were also discarded. The resulting structure dataset consists of 10173 (30% sequence identity), 14064 (50% sequence identity), and 17607 structures (90% sequence identity). To remove the bias from non-biological interface in the crystal asymmetric unit, the biological assembly provided by PDB was used. If multiple biological assemblies exist for one structure, the first assembly from PDB was used. For each of these structures, non-protein residues such as water, ion, and ligand were removed, and each protein residue and its $N$ closest ($N$=10, 15, 20, 25, 30, ranked based on $C_\alpha$-$C_\alpha$ distance) neighboring residues were extracted as a structural cluster. Clusters that have any atoms with an occupancy < 1 or missing backbone atoms were discarded. Protein oligomeric state was also considered during cluster extraction so that if a structure contains several identical subunits, only one of the subunits was used. Each cluster was then translated and orientated so that the $C_\alpha$, N, and C atoms of the target residue are located at the origin, the –$x$ axis, and the $z$=0 plane, respectively.

The input features for the neural networks are (1) for the central residues: *cos* and *sin* values of backbone dihedrals φ, ψ and ω, total solvent accessible surface area (SASA) of backbone atoms ($C_\alpha$, N, C, and O), and three-type (helix, sheet, loop) secondary structure; (2) for the neighbor residues: *cos* and *sin* values of backbone dihedrals φ, ψ, and ω, total SASA of backbone atoms, $C_\alpha$-$C_\alpha$ distance to the central residue, unit vector from the $C_\alpha$ atom of the central residue to the $C_\alpha$ atom of the neighbor residue, $C_\alpha$-N unit vector of the neighbor residue, $C_\alpha$-C unit vector of the neighbor residue, three-type secondary structure, and number of backbone-backbone hydrogen bonds between the central residue and the neighbor residue. The $C_\alpha$-$C_\alpha$ distance, $C_\alpha$-$C_\alpha$, $C_\alpha$-N, and $C_\alpha$-C unit vectors were used to define the exact position and orientation of the neighbor residue with respect to the central residue. *cos* and *sin* values of the dihedrals were used because the dihedrals that range from -180 to 180 are not continuous at -180 and 180. The SASA value was calculated using the Naccess program[67] on the whole protein structure (not on a structural cluster) with the sidechain atom removed, because during protein design, the identity of a residue and thus its sidechain atoms are unknown. Secondary structure was assigned with Stride.[68] All other features were calculated with an in-house program.

**Deep neural-network learning**
The neural network was constructed using the Keras library (http://keras.io) with rectified linear unit (ReLU) as the activation function for all layers. Training was performed using the categorical cross entropy as the loss function and the stochastic gradient descent method for optimization with a learning rate of 0.01, a Nesterov momentum of 0.9, and a batch size of 40,000. To account for the different abundance of each residue type in the training set, the training samples were weighted as: $W_i = N_{max}/N_i$, where $N_{max}$ is the maximal number of samples of all 20 residue types, and $N_i$ is the number of samples of residue type $i$. This bias would force the neural network to learn more from the residue types that are underrepresented in the training set. The output of the neural-network is the probability of 20 amino acids for the central residue of a cluster.

**Rosetta design**
Rosetta design was carried out with the *fixbb* program and talaris2014 score in Rosetta 3.7.[62] The crystal structures of the design targets were used as inputs without any prior minimization. 500 designs were performed for each protein with and without residue-type restraints, which were incorporated using the "-*resfile*" option.

## Acknowledgements
This work was supported by the National Natural Science Foundation of China (Grant no. 31700646) to Y.Q. and (Grant no. 21433004) J.Z., Ministry of Science and Technology of China (Grant no. 2016YFA0501700), NYU Global Seed Grant, and Shanghai Putuo District (Grant 2014-A-02) to J.Z.. We thank the Supercomputer Center of East China Normal University for providing us computer time.